\documentclass[aip,twocolumn,letterpaper]{revtex4}

\usepackage{fullpage}

\usepackage{graphics}
\usepackage{epsfig}

\usepackage[margin=1.0in]{geometry}

\begin{document}
\title{Magnetic-field properties in non-axisymmetric divertors }
\author{Allen H Boozer}
\affiliation{Columbia University, New York, NY  10027\\ ahb17@columbia.edu}

\begin{abstract}

Stellarator power plants require a plan for the removal of the particles and the heat that are exhausted across the plasma edge.  Unless a  flowing liquid metal can be used to carry the helium exhaust to places where it can be removed from the plasma chamber, the particle exhaust must be magnetically diverted into pumping chambers.  Studies are required to determine how magnetic features relate to the required divertor properties, how these magnetic features can be produced, and how they can be controlled.  General studies are clarified and simplified by the use of the magnetic field line Hamiltonian $\psi_p(\psi,\theta,\varphi)$ and a vector $\vec{x}(\psi,\theta,\varphi)$ that gives the point in space associated with each point in the $(\psi,\theta,\varphi)$ canonical coordinates, a flux and two angles.  The non-resonant Fourier terms in $\psi_p$ can be removed by a canonical transformation, so only resonant Fourier terms can determine the field line properties in the plasma edge and divertor.  This paper  discusses the important divertor properties and explains how $\psi_p(\psi,\theta,\varphi)$ and $\vec{x}(\psi,\theta,\varphi)$ can be obtained numerically in a special form for any stellarator magnetic field, $\vec{B}(\vec{x})$.  This form holds between an arbitrary magnetic surface and the chamber walls with the non-resonant terms eliminated.  Studies based on variations in the terms in such derived field-line Hamiltonians can determine what magnetic features are mathematically possible and how they could be produced and controlled by the external magnetic field coils.

\end{abstract}

\date{\today} 
\maketitle



\section{Introduction}

\subsection{Options for plasma exhaust}

In toroidal fusion systems, the plasma exhaust, which flows across the outermost confining magnetic surface, can (1) be allowed to neutralize by striking the chamber walls, which conventionally meant neutrals recycling into the plasma or (2) be directed to a divertor chamber to be pumped away. 

Two issues make the conventional recycling option unacceptable: (a) wall erosion and (b) in a deuterium-tritium power plant, the accumulation of the helium ash.  However, these two issues may be solvable by the use of flowing liquid lithium \cite{Andruczyk:2023}.  Wall damage, including that from high energy alpha particles, is relatively easily addressed by a thin layer of flowing lithium.  But, to be an alternative to a divertor, flowing liquid lithium must be able to trap cold alpha particles and carry them and other species in the escaping plasma to places where they can be removed from the plasma chamber.  As discussed in \cite{Andruczyk:2023}, adequate lithium trapping of helium was until recently considered unlikely but recent experiments imply the opposite may be true.  This could remove the requirement for a divertor system to concentrate the plasma flow to specific locations where pumps are located and fundamentally change the design of fusion power plants.



The pumping option comes with its own constraints, which are reviewed in \cite{Tok-Div2022},  \emph{Benefits and Challenges of Advanced Divertor Configurations in DEMO}.  (a) Practicality and the requirement that most of the volume immediately behind chamber walls be reserved for neutron breeding of tritium implies the divertor chambers can cover only a small fraction,  $f_d\sim0.01$, of the area of the walls.  (b) The pumping speed must be adequate to balance the turnover in the particle inventory in a particle confinement time $\tau_p$. 

The power crossing each square meter of the walls must be sold to pay for all of the equipment behind the walls---the higher the power density the more economic is fusion until the limits of materials are reached.  This point was made in \cite{Tok-Div2022}, the cost of fusion power is reduced by making the machine \emph{as small as possible for a target power output...One of the principle size-drivers...is the performance of the divertor in terms of the power which can be allowed to cross the separatrix.}  For any material surface there is a maximum thermal power density, $\sim10~$MW/m$^2$.  The flux of power in 14~MeV neutrons passing through the walls is four times that of heat in deuterium-tritium fusion and has a separate limit, the fluence $\sim10~$MW~yr/m$^2$.  The fluence is the power density multiplied by the time for damage due to the neutrons to accumulate.  The most economic power density is, therefore, dependent on how quickly in-vessel components can be replaced and how brittle in-vessel components can be allowed to become.  Stellarator coils can be designed to have open access to the plasma chamber \cite{Yamaguchi:2019}, which should greatly shorten the downtime required for the replacement of in-vessel components.  Stellarator plasmas can be robustly stable, which greatly reduces the danger of large transient loads.    A marginally economic fusion power plant has a total power density on the walls of $\sim1~$MW/m$^2$.  Even then, the smallness of $f_d$ makes the power loading of the divertor structures challenging unless most of the power can the radiated well before reaching the divertor \cite{Tok-Div2022}.

Without a strong back pressure along the diverted lines, the flow along the lines will reach the speed of sound, $C_s(T)\approx (7\times10^3~$m/s$)\sqrt{T/100\mbox{eV}}$.   The requirement that the pumping speed match the rate of particle loss from the plasma gives a simple condition 
\begin{equation} C_s(T_d) \approx \frac{n_p}{n_d}\frac{L_c/\tau_p}{f_d} . \label{Cs cond} \end{equation}  The diverted field lines have a connection length $L_c$ between the main plasma body and the divertor chamber.   The ratio $n_p/n_d$ is the ratio of the average plasma density to the divertor density.  Typical values in a power plant may be $n_p/n_d\sim10$, $f_d\sim0.01$, $L_c \sim 50~$m, and  $\tau_p\sim 10~$s, which gives the required temperature on the diverted field lines, $T_d\sim50~$eV.  

The pump itself can only remove neutral particles of density $n_n$ flowing into the pump at their speed of sound which is given by an even lower temperature $T_n\sim1~$eV.  This is consistent only when the neutral density at the pump is very high, not only compared to $n_d$ but also to $\bar{n}_n$ the average neutral density in the plasma chamber.  Otherwise, the plasma itself will be a better pump since it ionizes neutrals that strike it and has an area  $\approx1/f_d$ larger than the divertor chamber.   The ratio $n_n/\bar{n}_n$ is called the compression ratio and needs to be larger than approximately $1/f_d$.

In an axisymmetric tokamak, there is only one basic divertor concept---defined by an X-point and a separatrix.  Although there are design variants, double-null, snowflake, and super-X, the freedom to design around issues in an axisymmetric fusion power plant \cite{Tok-Div2022} is limited.  Non-axisymmetric divertor solutions can be used on tokamaks in two ways: (a) Coils that produce short wavelength magnetic fields can be installed next to the plasma edge that affect the edge but not the plasma core.  The feasibility of such coils in a power plant is questionable at best.  (b) Quasi-symmetry \cite{Boozer:1983} implies that axisymmetric-like neoclassical transport can be maintained by careful design while having large deviations from axisymmetry.  The computational and experimental demonstration that this technique can be used to obtain non-axisymmetric control of the plasma edge without affecting the core of a tokamak was given in \cite{Park:2018}.  The associated long wavelength magnetic fields are feasible in a power plant.

Stellarators offer many options for designing around problems associated with divertors---the very richness of possibilities forces a careful consideration of the research strategy.  

\subsection{Stellarator divertors}

The most refined and tested stellarator divertor concept is the island divertor \cite{W7-X_div2022}  of W7-X.  The computational tools developed for studying that divertor, such as EMC3-Eirene \cite{EMC3-Eirene}, and simplified models \cite{Feng:2022} provide tools and insights that are useful for a broad range of divertor types.  An important feature of the W7-X divertor is its long connection length $L_c$ compared to axisymmetric tokamak divertors, which is important for obtaining stable detachment.  When $L_c$ is too long for Equation (\ref{Cs cond}) to have a solution, divertor action is lost and the plasma exhaust diffuses to the surrounding walls.  Therefore, a detached divertor solution requires that $L_c$ neither be too short nor too long.

The plasma in a detached divertor recombines before striking a wall, which greatly eases the erosion and heat-load issues of tokamak divertors \cite{Tok-Div2022}.  However, stellarator island divertors place strong constraints on the plasma configuration: the edge rotational transform must be held to a low order rational number, such as $\iota=1$, during the plasma operation, and the shear $d\iota/dr$ must be small for two reasons: (1) The island should be of adequate width $\Delta_i$ to affect the plasma flow with a small perturbation.  (2) The rate at which a field encircles the island $(d\iota/dr)\Delta_i$ should be sufficiently small to give a long connection length.  


Non-resonant divertors naturally arise in stellarators, which have an outermost confining magnetic surface, and were the original idea for the W7-X divertor \cite{Strumbeger:divertor1992}.  Magnetic field line chaos arises just outside the outermost surface. 

Magnetic field line chaos is an exponentially increasing separation between neighboring lines with distance, but there are a number of subtleties.  These can be understood by studying the behavior of the magnetic field lines in the vicinity of an arbitrarily chosen line,  Appendix \ref{sec:line-chaos}.   For applications to both divertors and disruptions, chaos is better defined over a volume, which is discussed in other subsections of Appendix \ref{sec:chaos}.  Chaos cannot occur in an axisymmetric divertor and has little importance in the W7-X divertor, but it is more important in non-resonant divertors and could be made more important in designs for island divertors.  Magnetic field line chaos requires three spatial dimensions.  However, two-dimensional time-dependent flows are generically chaotic.   The beginning of Appendix A gives a brief discussion of everyday examples of stirring enhancing mixing, which can be thought of two-dimensionally, on the hope they increase intuition about the effect of chaos.

 Naively, one would expect chaotic field lines to have insufficient collimation and resilience to equilibrium changes for consideration as the basis for a divertor.  These expectations are not correct, as illustrated by numerical studies of the HSX magnetic field \cite{Bader-Boozer:2017,HSX divertor:2018}. This is due to transport barriers that naturally arise in chaotic field regions and provide the collimation, which can be studied in many ways, including simple maps \cite{Punjabi2022}.  It is important to understand what properties of the magnetic field determine the barriers and how the required externally-produced magnetic-field distributions are to be chosen and controlled.   

Non-resonant divertors have freedom that may have important implications beyond just removing island-divertor constraints on the rotational transform, $\iota$, and $d\iota/dr$.  For example, open field lines can go great distances in a chaotic region just outside the outermost confining surface with the parallel to cross-field transport of various types in principle controllable.  One could in principle use this chaotic region to produce a radiative mantel to reduce the plasma temperature from kilovolts at the last confining magnetic surface to the approximately a hundred electron-Volts, which is required to carry sufficient plasma to the divertor chamber.   Electromagnetic radiation from the mantel would spread the exiting heat more uniformly over the chamber walls than can any other method.   The chaotic region could also recycle main-chamber neutrals into the divertor channel, increasing the neutral compression ratio, and shield the main plasma from impurities.  


\subsection{Cross-field transport in chaotic regions}

Although EMC3-Eirene \cite{EMC3-Eirene} does not calculate the electric field in the divertor region that is required for quasi-neutrality, quasi-neutrality implies the electric potential along a divertor field line must have a variation $\Phi\sim T/e$.  A tube of divertor magnetic field lines that are being intermixed by the chaos has a characteristic width $\Delta$, the temperature difference between the lines in the tube is $T_d$.  The expected cross-field drift $\vec{E}\times\vec{B}/B^2 \sim T_d/(e\Delta B)$, and the characteristic diffusion coefficient is $D\sim\Delta^2/\tau_c$, where the characteristic time scale $\tau_c=\Delta/(\vec{E}\times\vec{B}/B^2)\sim \Delta^2 eB/T_d$.  Consequently, the diffusion coefficient has a Bohm-like form, $D\sim T_d/eB$.  The diffusion itself controls how large a temperature difference can arise over the scale of the chaotic intermixing as the plasma flows along the field lines at the speed of sound $C_s$ over a connection length $L_c$.  That is,  $\sqrt{DL_c/C_s}\sim \Delta$.  When $T=100~$eV, $B=5$~T, and the connection length $L_c= 50$~m, one finds that $T_d/T\sim (\Delta/0.4~\mbox{m})^2$.   In other words, the natural equilibrium in a chaotic field diffuses particles to give quasi-neutrality over the flux tubes in which field lines are chaotic and thereby intermix.

The natural diffusive intermixing given by field-line chaos is in addition to any diffusive microturbulent transport, which is characteristically gyro-Bohm like.  Mictroturblent transport is due to displacements of the plasma by a distance $\xi$ across the lines.  Using an analogous argument to the one in the previous paragraph, the drift is $\vec{E}\times\vec{B}/B^2 \sim T_d/(e\xi B)$, where $T_d$ is the temperature difference in the plasma on the spatial scale $\xi$.  This implies $T_d\sim\xi/a$, where $a$ is the thermal scale length, which in the plasma interior can be approximated by the minor radius.  The same argument as used for chaotic field lines gives at turbulent transport coefficient $D_t\sim \xi^2/\tau_c \sim (\xi/a)T/eB$.  Gyro-Bohm transport is obtained when the typical displacement produced by the microtubulent fluctuations is $\xi\sim \rho_s$, where $\rho_s$ is the ion gyroradius calculated with the speed of sound; $\rho_s\equiv m_iC_s/eB$.

\subsection{Overview of paper}

What magnetic field properties are physically possible and controllable need to be investigated as well their broader implications for the design of stellarators that offer a low risk and fast development path for fusion power plants.


This paper develops a method based on the magnetic field line Hamiltonian $\psi_p(\psi,\theta,\varphi)$ for studying what is possible and what is controllable about the magnetic field line properties in the plasma edge and divertor of an arbitrary non-axsiymmetric toroidal plasma, such as a stellarator or a perturbed tokamak.  This approach allows general studies of the plasma edge and divertor region that would not otherwise be possible.  It gives insights on what external magnetic fields are needed for divertor optimization and control.

As shown in Section \ref{sec:determination Psi_p}, a given magnetic field can be separated into two parts.  (1) A Hamiltonian in canonical coordinates, which is the poloidal magnetic flux $\psi_p$ as function of the toroidal magnetic flux $\psi$, a poloidal angle $\theta$, and a toroidal angle $\varphi$.  (2) The position $\vec{x}(\psi,\theta,\varphi)$ in an ordinary set of spatial coordinates, Cartesian or cylindrical, of each point in the canonical coordinates, $(\psi,\theta,\varphi)$, of the Hamiltonian.  

Define $\iota(\psi)\equiv d\big<\psi_p\big>/d\psi$, where $\big<\psi_p\big>$ is $\psi_p$ averaged over a constant-$\psi$ surface.  Then, Section \ref{sec:simplification} shows that when $\psi_p(\psi,\theta,\varphi)$ is Fourier decomposed any term that does not resonate  with the transform $\iota(\psi)$ can be removed from the Hamiltonian by a canonical transformation. A canonical transformation includes a change in the function $\vec{x}(\psi,\theta,\varphi)$ so the function $\vec{B}(\vec{x})$ is unchanged. The method given in Sections \ref{sec:determination Psi_p} and \ref{sec:simplification} is related to but distinct from the method and examples \cite{Kuo-Petravic:1987} given by Kuo-Petravic and Boozer in 1987. 

Once the non-resonant Fourier terms in $\psi_p$ are removed, the general Hamiltonian for the magnetic field in the edge and divertor region consists of two terms $\psi_p(\psi,\theta,\varphi)=\big<\psi_p\big>(\psi) +\hat{\psi}_p(\psi,\theta,\varphi)$, where every $m,n$ term in the Fourier series for $\hat{\psi}_p$ is resonant, which means $n/m=\iota(\psi)\equiv d<\psi_p>/d\psi$ somewhere in the spatial region of interest.

The magnetic properties of the edge and divertor in an arbitrary stellarator configuration are determined by $\iota(\psi)$ and the resonant Fourier terms in $\hat{\psi}_p$.  As discussed in Section \ref{Discussion}, these can be varied to study their importance to the properties of the divertor.  

While changing $\hat{\psi}_p$, small changes to the position vector are required to maintain the correct current density, for example $\vec{j}=0$.  In many cases, the required position-vector changes would appear to be negligibly small, Appendix \ref{shape-j}.   The external, coil-produced, magnetic field would need to be changed, and these changes define the method of control.  It is relatively simple to determine what external magnetic fields are required and how these fields affect the overall plasma optimization.  

Since $\iota(\psi)$ determines what Fourier terms in $\psi_p$ are resonant, it is particularly important to study the robustness of divertor configurations to changes in $\iota(\psi)$ and its shear and how the effects of those changes can be compensated.

Appendix \ref{sec:chaos} is on magnetic field line chaos, which is an important concept not only for non-axisymmetric divertors but also for magnetic reconnection.  This includes the fast breaking of magnetic surfaces in tokamak disruptions \cite{Boozer:surfaces,Jardin:2022} and more generally sudden changes in magnetic field line connections when a magnetic field is evolving \cite{Boozer:B-ev}.


\section{Determining a Field-line Hamiltonian \label{sec:determination Psi_p}}

To study the effect of changes in the magnetic field line Hamiltonian, one must first find the Hamiltonian in a convenient coordinate system for the plasma edge.  A convenient choice is $\vec{x}(\rho,\theta,\varphi)$, which is an interpolation between an outer magnetic surface at $\rho=0$ and the wall at $\rho=1$.  In $(R,\varphi,Z)$ cylindrical coordinates,
\begin{eqnarray}
&&\vec{x}(\rho,\Theta,\varphi)= R(\rho,\Theta,\varphi)\hat{R}(\varphi) + Z(\rho,\Theta,\varphi)\hat{Z}; \label{x-vector} \\
&& R = R_a(\varphi) + r(\rho,\Theta,\varphi) \cos\Theta \hspace{0.2in} \mbox{and   } \\ 
 &&Z = Z_a(\varphi) -r\sin\Theta \hspace{0.2in}\mbox{  where } \\
 &&r = r_m(\Theta,\varphi)(1-\rho) + r_w(\Theta,\varphi) \rho. \label{r exp}
\end{eqnarray} 

The magnetic field can be written in terms of two fluxes, $\Psi$ a toroidal flux and $\Psi_p$ a poloidal flux \cite{Boozer:RMP},
\begin{equation}
2\pi\vec{B}=\vec{\nabla}\Psi(\rho,\Theta,\varphi)\times \nabla\Theta + \vec{\nabla}\varphi\times\vec{\nabla}\Psi_p(\rho,\Theta,\varphi), \hspace{0.2in}
\end{equation}

The theory of general coordinates, which is explained in a two-page appendix to Reference \cite{Boozer:RMP} implies
\begin{eqnarray}
\frac{\partial\Psi}{\partial\rho} &=&  2\pi \vec{B}\cdot\left(\frac{\partial \vec{x}}{\partial\rho}\times \frac{\partial \vec{x}}{\partial\Theta}\right) \\
&\equiv&F_\Psi(\rho,\Theta,\varphi) \hspace{0.2in} \mbox{   and   } \label{Psi-rho}\\
\frac{\partial\Psi_p}{\partial\rho} &=& 2\pi \vec{B}\cdot\left(\frac{\partial \vec{x}}{\partial\varphi}\times \frac{\partial \vec{x}}{\partial\rho}\right)\\
&\equiv& F_p(\rho,\Theta,\varphi) \label{Psi_p-rho}
\end{eqnarray}

Since $\rho=0$ is a magnetic surface, the toroidal flux is a constant there, $\Psi_0$.  By picking a regular array of $\Theta$ and $\varphi$ points, an ordinary differential equation,
\begin{equation}
\frac{d\rho}{d\Psi} = \frac{1}{F_\Psi(\rho,\Theta,\varphi)} \label{d rho / d Psi},
\end{equation}
can be integrated for each $(\Theta,\varphi)$ point from $\rho=0$ at $\Psi=\Psi_0$ to obtain $\rho(\Psi,\Theta,\varphi)$.  A Fast Fourier Transform then determines the coefficients in the Fourier series, which has only cosinusoidal terms when the field is stellarator symmetric;
\begin{equation}
\rho(\Psi,\Theta,\varphi) = \sum_{mn} \rho_{mn}(\Psi) \cos(m\Theta-n\varphi).
\end{equation}
For an analytic magnetic field, the Fourier amplitudes converge exponentially.

Similarly Equations (\ref{Psi_p-rho}) and (\ref{Psi-rho}) imply that
\begin{equation}
\frac{d\Psi_p}{d\Psi} = \frac{F_p(\rho,\Theta,\varphi)}{F_\Psi(\rho,\Theta,\varphi)}, 
\end{equation}
which can be integrated for each $(\Theta,\varphi)$ point from $\Psi_p(0)$ at $\Psi=\Psi_0$, while simultaneously integrating Equation (\ref{d rho / d Psi}),  to obtain $\Psi_p(\Psi,\Theta,\varphi)$.  Another Fast Fourier Transform then determines the coefficients in a Fourier series, which has only cosinusoidal terms when the field is stellarator symmetric;
\begin{equation}
\Psi_p(\Psi,\Theta,\varphi) = \sum_{mn} \Psi^p_{mn}(\Psi) \cos(m\Theta-n\varphi). \label{orig-B-line-Hamiltonian}
\end{equation}

$\Psi_p(\Psi,\Theta,\varphi)$ of Equation (\ref{orig-B-line-Hamiltonian}) is the Hamiltonian for the magnetic field lines expressed as a function of its canonical coordinates;
\begin{eqnarray}
&&\frac{d\Psi}{d\varphi} = - \frac{\partial \Psi_p}{\partial\Theta};\\
&&\frac{d\Theta}{d\varphi} = \frac{\partial \Psi_p}{\partial\Psi}.
\end{eqnarray}

The function $\rho(\Psi,\Theta,\varphi)$ through Equations (\ref{x-vector}) to (\ref{r exp}) gives to spatial location of each point of the canonical coordinates $(\Psi,\Theta,\varphi)$.


\section{Hamiltonian simplification \label{sec:simplification}}

For $\Psi$ just slightly larger than $\Psi_0$, the field-line Hamiltonian has the form $\Psi_p=\big<\Psi_p\big>(\Psi)$ plus small Fourier terms $\Psi_{mn}^p\cos(m\Theta-n\varphi)$, where $\big<\cdots\big>$ is an average over the poloidal and toroidal angles.  These terms are of two types: resonant terms with $n/m = \iota$, where $\iota\equiv d\big<\Psi_p\big>/d\Psi$, and non-resonant terms $n/m \neq \iota$ in the region of interest.  Canonical transformations of the canonical coordinates can be used to simplify the Hamiltonian by removing the non-resonant terms from $\Psi(\Psi,\Theta,\varphi)$.  

Canonical transformations change both the magnetic field line Hamiltonian $\Psi_p(\Psi,\Theta,\varphi)$ and the dependence of the spatial vector on the canonical coordinates, $\vec{x}(\Psi,\Theta,\varphi)$, so $\vec{B}(\vec{x})$ remains the same.  Consequently, magnetic field line trajectories $\vec{x}(\varphi)$ remain identical when the canonical coordinates are transformed from their old form $(\Psi,\Theta,\varphi)$ to a new $(\psi,\theta,\varphi)$.  For the divertor problem, this transformation can be done by a series of infinitesimal canonical transformations.

 
\subsection{Infinitesimal canonical transformations}

The full canonical transformation equations between the magnetic field line Hamiltonian $\Psi_p(\Psi,\Theta,\varphi)$ and $\psi_p(\psi,\theta,\varphi)$ are defined by a generating function $S(\theta,\Psi,\varphi)$,
\begin{eqnarray}
\psi &=& \frac{\partial S(\theta,\Psi,\varphi)}{\partial\theta}\\
\Theta &=& \frac{\partial S}{\partial\Psi}\\
\Psi_p(\Psi,\Theta,\varphi) &=& \psi_p(\psi,\theta,\varphi)+ \frac{\partial S}{\partial\varphi}
\end{eqnarray}

This transformation can be accomplished by a series of infinitesimal canonical transformations.  Let $S(\theta,\Psi,\varphi)=\Psi\theta - \epsilon s$, then $\psi = \Psi- \epsilon \partial s/\partial\theta$ and $\Theta = \theta -\epsilon \partial s/\partial\Psi$.  As $\epsilon\rightarrow0$,
\begin{eqnarray}
\left(\frac{\partial \psi}{\partial\epsilon}\right)_{\Psi\Theta\varphi} &=& -\frac{\partial s(\psi,\theta,\varphi)}{\partial\theta} \\
\left(\frac{d \theta}{\partial \epsilon}\right)_{\Psi\Theta\varphi} &=&  \frac{\partial s}{\partial\psi}\\
\left(\frac{\partial  \psi_p}{\partial \epsilon}\right)_{\Psi\Theta\varphi} &=&  \frac{\partial s}{\partial\varphi} .\label{Ham-change}
\end{eqnarray}

When a quantity is changed from the dependence $f(\Psi,\Theta,\varphi,\epsilon)$ to a dependence $f(\psi,\theta,\varphi,\epsilon)$,
\begin{eqnarray}
\left(\frac{\partial  f}{\partial \epsilon}\right)_{\Psi\Theta\varphi} &=& \left(\frac{\partial  f}{\partial \epsilon}\right)_{\psi\theta\varphi} + \left(\frac{\partial  f}{\partial \psi}\right)_{\theta\varphi}\left(\frac{\partial  \psi}{\partial \epsilon}\right)_{\Psi\Theta\varphi} \nonumber\\
&& +\left(\frac{\partial  f}{\partial \theta}\right)_{\psi\varphi}\left(\frac{\partial  \theta}{\partial \epsilon}\right)_{\Psi\Theta\varphi}\\
&=&\frac{\partial  f(\psi,\theta,\varphi,\epsilon)}{\partial \epsilon} + [f,s],  \label{infin.canon}
\end{eqnarray}
where the Poisson bracket is defined by 
\begin{equation} [f,s]\equiv \frac{\partial f}{\partial\theta}\frac{\partial s}{\partial\psi}-\frac{\partial f}{\partial\psi}\frac{\partial s}{\partial\theta}.\end{equation}
Equation (\ref{infin.canon}) gives the change in $f$ due to a change in the coordinates, not an intrinsic change, when $(\partial f/\partial\epsilon)_{\Psi,\Theta,\varphi}=0$.  

Equation (\ref{Ham-change}) together with Equation (\ref{infin.canon}) imply that without an intrinsic change in the magnetic field
\begin{eqnarray}
\frac{\partial  \psi_p(\psi,\theta,\varphi,\epsilon)}{\partial \epsilon}&=& \frac{\partial s}{\partial\varphi}-[\psi_p,s]. \label{d psi}
\end{eqnarray} 
The change in the coordinates due to the canonical transformation is
\begin{eqnarray}
&&  \frac{\partial \vec{x}}{\partial \epsilon}=-[\vec{x},s], \hspace{0.1in} \mbox{which is equivalent to   } \label{d x}\\
&& \frac{\partial \rho}{\partial \epsilon}=-[\rho,s] \hspace{0.1in} \mbox{and   }\\
&& \frac{\partial \Theta}{\partial \epsilon}=-[\Theta,s].
\end{eqnarray}
At $\epsilon=0$, the Hamiltonian is $\psi_p(\psi,\theta,\varphi)=\Psi_p(\Psi,\Theta,\varphi)$,  $\psi=\Psi$, and $\theta=\Theta$.   

Equations such as (\ref{d psi}) and (\ref{d x}) can be interpreted as ordinary differential equations in $\epsilon$ for each value of $(\psi,\theta,\varphi)$.  After each infinitesimal step in $\epsilon$, a new infinitesimal canonical transformation can be performed, which is equivalent to making $s$ depend on $\epsilon$.  Since any series of canonical transformations is itself an exact canonical transformation, one can choose the function $s(\psi,\theta,\varphi,\epsilon)$ so the desired answer is obtained at $\epsilon=1$.  The functions $\vec{x}(\psi,\theta,\varphi)$ and $\psi_p(\psi,\theta,\varphi)$ give exactly the same magnetic field line trajectories $\vec{x}(\varphi)$ as as the original functions $\vec{x}(\Psi,\Theta,\varphi)$ and $\Psi_p(\Psi,\Theta,\varphi)$.


\subsection{Elimination of non-resonant terms in $\psi_p$}

At $\epsilon=0$, the Hamiltonian $\psi^{(0)}_p(\psi,\theta,\varphi)$ has the form 
\begin{equation}
\psi^{(0)}_p = \big<\psi^{(0)}_p\big>(\psi) + \tilde{\psi}^{(0)}_p(\psi,\theta,\varphi)+\hat{\psi}^{(0}_p(\psi,\theta,\varphi),
\end{equation} 
where every term in a Fourier decomposition of $\hat{\psi}^{(0)}_p$ has a resonance somewhere in the region of interest, which means a place where a term $\cos(m\theta-n\varphi)$ satisfies $n/m =\iota(\psi)\equiv d\big<\psi^{(0)}_p\big>/d\psi$, but no Fourier terms in $\tilde{\psi}^{(0)}_p$ are resonant.  Both $\tilde{\psi}^{(0)}_p$ and $\hat{\psi}^{(0)}_p$ when integrated over $\theta$ and $\varphi$ are zero.  

The choice $\tilde{\psi}_p=(1-\epsilon)\tilde{\psi}^{(0)}_p$ ensures that when $\epsilon=1$ the Hamiltonian consists of  a term dependent on $\psi$ alone plus a term that is resonant somewhere in the region of interest.
Equation (\ref{d psi}) implies
\begin{eqnarray}
\frac{\partial \big<\psi_p\big>}{\partial\epsilon} - \tilde{\psi}^{(0)}_p +\frac{\partial \hat{\psi}_p}{\partial\epsilon} &=& \frac{\partial s}{\partial\varphi}-[\big<\psi_p\big>,s]\nonumber\\&&-(1-\epsilon)[\tilde{\psi}^{(0)}_p,s] - [\hat{\psi}_p,s]. \nonumber\\
\end{eqnarray}
This equation can be written as three separate equations for the three parts of $\psi_p$.  
\begin{eqnarray}
&&\frac{\partial \big<\psi_p\big>}{\partial\epsilon} =-(1-\epsilon) \Big<[\tilde{\psi}^{(0)}_p,s]\Big> - \Big<[\hat{\psi}^{(0)}_p,s]\Big>; \label{<psi_p>} \\
&&\frac{\partial \hat{\psi}_p}{\partial\epsilon} =-[\hat{\psi}_p,s] +\Big<[\tilde{\psi}_p,s]\Big> +(1-\epsilon)\sigma_r -\sigma_{nr}; \label{res-eq} \nonumber\\ \\
&&\frac{\partial s}{\partial\varphi}+\iota(\psi,\epsilon)
\frac{\partial s}{\partial\theta}= - \tilde{\psi}^{(0)}_p \nonumber\\ &&\hspace{0.2in} +(1-\epsilon)\Big([\tilde{\psi}^{(0)}_p,s] -\Big<[\tilde{\psi}^{(0)}_p,s]\Big>  - \sigma_{r} \Big) +\sigma_{nr}, \label{s-eq} \nonumber\\
\end{eqnarray} 
where each Fourier term in $\sigma_r$ is resonant somewhere in the region of interest and $\sigma_{nr}$ has no resonant Fourier terms.  The function $\sigma_r$ is chosen so the right-hand side of Equation (\ref{s-eq}) has no resonant terms, which means this equation has a unique solution for the periodic function $s$.  Note that   $[\big<\psi_p\big>,s]=-\iota(\psi,\epsilon)(\partial s/\partial\theta)$.  The function $s_{nr}$ is chosen so non-resonant terms are removed from Equation (\ref{res-eq}) for $\partial \hat{\psi}_p/\partial\epsilon$.  

\subsection{Iterative Solution}

A simpler method of removing the non-resonant terms than solving Equations (\ref{<psi_p>})  through (\ref{s-eq}) is to proceed iteratively.  Any function $s(\psi,\theta,\varphi,\epsilon)$ defines an infinitesimal canonical transformation, and  a succession of infinitesimal transformations is a canonical transformation.   

Equation (\ref{s-eq}) suggests an $\epsilon$ independent form for the infinitesimal generating function for an iterative solution:
\begin{eqnarray}
s_I(\psi,\theta,\varphi) &=& \sum s_{mn}(\psi)\sin(m\theta-n\varphi); \\
\frac{\partial s_I}{\partial\varphi}+\iota_0(\psi)\frac{\partial s_I}{\partial\theta}&\equiv& - \tilde{\psi}^{(0)}_p \label{def-s_I} \\ &=&-\sum \tilde{\psi}^{p(0)}_{mn}(\psi)\cos(m\theta-n\varphi); \hspace{0.2in}\\
s_{mn}(\psi) &=&\frac{ \tilde{\psi}^{p(0)}_{mn}(\psi) }{n - \iota_0(\psi)m},
\end{eqnarray}
where $\iota_0$ is given by $<\psi_p^{(0)}>$.  The non-resonant part of the poloidal flux $\tilde{\psi}_p^{(0)}$ has a simple definition.  It contains all the Fourier terms in $\psi_p^{(0)}$ for which $s_{mn}$ is not singular.

To carry out the canonical transformation, define $\delta\psi_p$ so
\begin{eqnarray}
\psi_p &=& \big<\psi_p^{(0)}\big> + \tilde{\psi}_p^{(0)} + \delta\psi_p, \label{delta-psi_p}
\end{eqnarray}
which implies $\delta\psi_p^{(0)}=\hat{\psi}_p^{(0)}$ at $\epsilon=0$.  Equation (\ref{d psi}) implies
\begin{eqnarray}
\frac{\partial \psi_p}{\partial \epsilon} &=& \frac{\partial s_I}{\partial \varphi}+\iota_0 \frac{\partial s_I}{\partial \theta} - [(\psi_p-\big<\psi_p^{(0)}\big>),s_I] \hspace{0.1in} \\
&=& - \tilde{\psi}_p^{(0)}  - [\tilde{\psi}_p^{(0)},s_I] - [\delta\psi_p,s_I].
\end{eqnarray}
Equation (\ref{delta-psi_p}) implies $\partial \delta\psi_p/\partial \epsilon= \partial \psi_p/\partial \epsilon$, so
\begin{eqnarray}
\frac{\partial \delta\psi_p}{\partial \epsilon} +[\delta\psi_p,s_I] &=&  - \Big( \tilde{\psi}_p^{(0)} + [\tilde{\psi}_p^{(0)},s_I] \Big),
\end{eqnarray}
which is a linear inhomogeneous equation for $\delta\psi_p$.  The inhomogeneity $\tilde{\psi}_p^{0} + [\tilde{\psi}_p^{0},s_I]$ is independent of $\epsilon$.  This equation is to be integrated from $\epsilon=0$ to $\epsilon=1$.  At $\epsilon=0$, $\delta\psi_p = \hat{\psi}_p$, the resonant part of the original magnetic field line Hamiltonian.  At $\epsilon=1$, which means after the canonical transformation is completed,
\begin{eqnarray}
\psi_p^{(1)} &=& \big<\psi_p^{(0)}\big>  + \tilde{\psi}_p^{(0)}+ \delta\psi_p^{(1)}.
\end{eqnarray}
The position function is transformed by
\begin{equation}
\frac{\partial\vec{x}}{\partial\epsilon} = - [\vec{x},s_I].
\end{equation}

If necessary, the $\theta,\varphi$ dependent part of the $\epsilon=1$ Hamiltonian $\psi_p^{1}$ can be separated into its resonant and non-resonant parts and iterated again to reduce the non-resonant Fourier terms further.

 
 \section{Discussion of divertor and edge studies \label{Discussion}}
 
 Given any stellarator magnetic field $\vec{B}(\vec{x})$, the magnetic field line Hamiltonian can be determined by the method given in Section \ref{sec:determination Psi_p}.  This Hamiltonain can be canonically transformed so all non-resonant Fourier terms in the variation of the Hamiltonian on the $\psi$ surfaces are removed using the method outlined in Section \ref{sec:simplification}.  The canonically transformed Hamiltonian is $\psi_p(\psi,\theta,\varphi)=\big<\psi_p\big>(\psi) + \hat{\psi_p}(\psi,\theta,\varphi)$, where every Fourier term in $\hat{\psi_p}$ resonates with the rotational transform $\iota(\psi)\equiv d\big<\psi_p\big>/d\psi$ at some value of $\psi$ in the region of interest.  The calculations also determine the canonically transformed position vector $\vec{x}(\psi,\theta,\varphi)$, so the function $\vec{B}(\vec{x})$ is unchanged.  The vector $\vec{x}(\psi,\theta,\varphi)$ gives the location in ordinary spatial coordinates of each point in $(\psi,\theta,\varphi)$ canonical coordinates.

It is the resonant Fourier terms in the magnetic field line Hamiltonian that determine the location in $\psi$ of the outermost confining magnetic surface and the topological behavior of the magnetic field lines outside that surface.  By changing these Fourier terms, studies can be made of what topological forms are possible and how they can be controlled.  
 
 Questions that need to be studied about the non-resonant Fourier terms in the plasma edge and divertor Hamiltonian include:
 \begin{itemize}
 \item  Which terms control the location of the strike points on the wall?  How wide or narrow can the stripes on the walls be made?
 \item  Which terms control the width of the diverted field line region near the plasma?  
 
 Magnetic field lines could make many toroidal transits in a chaotic region just outside the confining magnetic surfaces before leaving the vicinity of the plasma and striking the walls.  Such a region could be important for forming a radiative mantel and for shielding the plasma from of neutrals.  A radiative mantel may be needed for evenly spreading the heat over the walls even if a thin film of liquid lithium on the walls can be used to remove the helium and other species from the plasma chamber.
 \item  What is the sensitivity of important effects to $\iota(\psi)\equiv d\Big<\psi_p\Big>/d\psi$ and $d\iota/d\psi$ in the divertor region?
 \item  Can the most important terms be controlled by external magnetic field distributions while maintaining the physics quality of the core plasma?
 \end{itemize}
 
 When terms in $\psi_p$ are changed, the local current density $\vec{j}$ would be changed unless the function $\vec{x}(\psi,\theta,\varphi)$ is also changed.  But, when the changes in $\psi_p$ are small, the required change in $\vec{x}(\psi,\theta,\varphi)$ to keep a realistic current density, in many cases $\vec{j}\approx0$, is small, Appendix \ref{shape-j}.
 
\section*{Acknowledgements}

This material is based upon work supported by the grant 601958 within the Simons Foundation collaboration ``\emph{Hidden Symmetries and Fusion Energy}" and by the U.S. Department of Energy, Office of Science, Office of Fusion Energy Sciences under Award DE-FG02-95ER54333.

\section*{Author Declarations}

The author has no conflicts to disclose.


\section*{Data availability statement}

Data sharing is not applicable to this article as no new data were created or analyzed in this study.

 

 \appendix


\section{Magnetic field line chaos \label{sec:chaos} }

Magnetic field line chaos is important not only in non-axisymmetric divertors but also for allowing rapid reconnection in evolving near-ideal plasmas.  However, the source of the chaos differs between the reconnection and the divertor applications.

  In an ideal evolution, the poloidal flux can be assumed to be independent of time \cite{Boozer:RMP} with the evolution given by the time dependence of the position vector, $\vec{x}(\psi,\theta,\varphi,t)$.  The flow velocity of the canonical coordinates, $\vec{u}\equiv\partial\vec{x}(\psi,\theta,\varphi,t)/\partial t$, is, like other three-dimensional flows, generically chaotic.  This causes tubes of magnetic flux to become exponentially contorted, and the resistivity can interdiffuse field lines from different tubes in places where the tubes become exponentially thin \cite{Boozer:B-ev}.  This accounts for the ease with which magnetic surfaces can be broken \cite{Boozer:surfaces}, as in a disruption, no matter how small the resistivity may be. It should be noted, however, that the re-formation of magnetic surfaces in tokamaks after a disruption has no analogous enhancement from an ideal-evolution effect. 

In non-axisymmetric divertors, the chaos is not from the function $\vec{x}(\psi,\theta,\varphi)$ but from the Hamiltonian itself.  This appendix shows how chaos can be defined field line by field line, the volumetric definition of chaos, and how the chaos given by $\psi_p$ can be calculated.  The contrasting applications of chaos to disruptions and divertors are discussed. 

Chaotic magnetic fields require three dimensions and understandable figures are difficult to draw.  However, even two-dimensional time-dependent flows are generally chaotic.  The effect of stirring on enhancing mixing in fluids is apparent in everyday examples, which can be thought of two dimensionally, but the correct interpretation was not given until 1984 in a famous paper by Hassan Aref  \cite{Aref:1984}.  A familiar example is mixing a can of paint by stirring with a paddle---going around and around inside the can.  The colorant and the carrier start out separated.  At first the stirring causes the colorant to form longer but thinner streamers in the carrier.  The area of a streamer is preserved, and there is no intermixing.  The length to the width of the streamers increases exponentially with the number of stirs until the streamers become so thin that the weak interdiffusion between colorant and the carrier causes the separation between the two to disappear, rather suddenly, after of order ten stirs.   A similar effect allows a radiator to heat a room in approximately ten minutes rather than approximately a month that would be expected from the thermal conductivity of air.


\subsection{Individual field line definition of chaos \label{sec:line-chaos}}

Magnetic-field-line chaos means neighboring magnetic field lines exponentiate apart.  It is a subtle concept, but the subtleties can be understood field line by field line.

The behavior of magnetic field lines sufficiently close any magnetic field line that does not pass near a $\vec{B}=0$ point can be described exactly \cite{Boozer:separation}  by the expression $2\pi \vec{B} = \vec{\nabla}\tilde{\psi}\times\vec{\nabla}\alpha +\vec{\nabla}\ell \times\vec{\nabla}\tilde{H}$.  Here $\tilde{\psi}\rightarrow0$ is the magnetic flux in a tube around the chosen line, $\alpha$ is an angle about that flux tube, $\ell$ is the distance along the chosen line, and $\tilde{H}$ is the Hamiltonian for field lines in the neighborhood of the chosen line with $\tilde{H} =\tilde{\psi} h(\alpha,\ell)$ and $h=k_\omega( \ell) + k_q(\ell) \cos(2\alpha-2\phi_q(\ell))$.  The magnitude of the chaos along the chosen line is given by $2\sigma(\ell,\alpha_0)\equiv \ln(\tilde{\psi}/\tilde{\psi}_0)$ with $(\tilde{\psi}_0,\alpha_0)$ the $\ell=0$ position of the line.   Three functions of $\ell$ determine the magnitude of the chaos along the line: $k_\omega(\ell)$ is given by the torsion and the current density along the line, $k_q(\ell)$ is given by the amplitude of the quadrupole component of the magnetic field in an expansion around the line, and $\phi_q(\ell)$ is the phase of the quadrupole field.  

Chaos is important when the variation in the exponentiation $\tilde{\sigma}(\alpha_0)\equiv \sigma_{max}-\sigma_{min}$ associated with each line is far larger than unity for some $\alpha_0$ for field lines  throughout a volume.  An X-point in an axisymmetric field is not chaotic in an important sense for only field lines that lie exponentially close to the line that passes through the X-point exponentially separate.  In a non-axisymmetric field, field lines throughout a volume, whether bounded or not, can exponentiate apart by an arbitrarily large amount.

A magnetic field line is chaotic when its Lyapunov exponent, 
\begin{equation}\lambda_L(\alpha_0) \equiv \lim_{\ell\rightarrow\infty} \frac{\sigma(\ell,\alpha_0)}{\ell}, \end{equation}
is non-zero for some $\alpha_0$, but that is not a necessary condition for the importance of chaos.  The exponentiation $\tilde{\sigma}(\alpha_0)$ can be arbitrarily large for a field line with zero Lyapunov exponent.  An important example is given by the field lines in a magnetic surface \cite{Boozer:surfaces} that has become highly contorted by an ideal magnetic perturbation.  When $\tilde{\sigma}(\alpha_0)$ becomes larger than the natural logarithm of the ratio of the resistive timescale divided by the ideal evolution time, the breakup of the magnetic surface is unavoidable \cite{Boozer:B-ev}.


\subsection{Volumetric definition of chaos}

A definition of magnetic field line chaos that applies over a volume \cite{Boozer-Elder:2021}  is more useful for both divertor and disruption studies than the field line by field line definition.

A magnetic field line is given by $\vec{x}(\vec{x}_0,\varphi)$, where $\vec{x}_0$ is the point the field line passed through at $\varphi=0$.  Using the theory of general coordinates from the appendix of \cite{Boozer:RMP}, the equation for a magnetic field line is 
\begin{eqnarray}
\frac{d\vec{x}}{d\varphi} &=& \vec{h}(\vec{x}), \hspace{0.2in}\mbox{where   }\\
\vec{h}&\equiv& \frac{\vec{B}}{\vec{B}\cdot\vec{\nabla}\varphi}\\
&=& \frac{\partial \vec{x}(\psi,\theta,\varphi)}{\partial\varphi} + \frac{\partial\psi_p}{\partial\psi} \frac{\partial \vec{x}}{\partial\theta} - \frac{\partial\psi_p}{\partial\theta} \frac{\partial \vec{x}}{\partial\psi};\\
\frac{d\vec{x}}{d\varphi} &=& \frac{d\psi}{d\varphi}\frac{\partial \vec{x}}{\partial\psi}+\frac{d\theta}{d\varphi}\frac{\partial \vec{x}}{\partial\theta} + \frac{d\varphi}{d\varphi}\frac{\partial \vec{x}}{\partial\varphi}
\end{eqnarray} 
with $d\varphi/d\varphi=1$.

Chaos measures the separation between two magnetic field lines, lines that pass through $\vec{x}_0$ and $\vec{x}_0+\vec{\delta}_0$ in the limit as $\big| \vec{\delta}_0 \big|\rightarrow0$.  The equation obeyed by $\vec{\delta}$ is 
\begin{equation}
\frac{d\vec{\delta}}{d\varphi} = \vec{\delta}\cdot\vec{\nabla}\vec{h}.
\end{equation}  
This equation is linear, so the magnitude  $\big| \vec{\delta} \big|$ is irrelevant, but what is relevant is the direction of $\vec{\delta}_0$ across the field line at  $\varphi=0$.  There are two possible directions, which define two independent solutions.  These directions define corrdinates $\xi$ and $\eta$, and the two distances from the line are $\delta_\xi(\varphi)$ and $\delta_\eta(\varphi)$.   One of the two solutions is $\delta_\xi^{(\xi)}(\varphi)$ and $\delta_\eta^{(\xi)}(\varphi)$ with  $\delta_\xi^{(\xi)}(0)=1$ and $\delta_\eta^{(\xi)}(0)=0$.  The other solution is  $\delta_\xi^{(\eta)}(\varphi)$ and $\delta_\eta^{(\eta)}(\varphi)$ with  $\delta_\xi^{(\eta)}(0)=0$ and $\delta_\eta^{(\eta)}(0)=1$. 

The Jacobian matrix for the Lagrangian position vector associated with the position vector $\vec{x}(\vec{x}_0,\varphi)$ of all magnetic field lines in a region of space  is 
\begin{eqnarray}
\frac{\partial\vec{x}}{\partial\vec{x}_0} &\equiv& \left(\begin{array}{cc}\frac{\partial \xi}{\partial \xi_0} & \frac{\partial \xi}{\partial \eta_0} \\\frac{\partial \eta}{\partial \xi_0} & \frac{\partial \eta}{\partial \eta_0}\end{array}\right), \mbox{   so  } \\
&=&\left(\begin{array}{cc} \delta_\xi^{(\xi)} & \delta_\eta^{(\xi)}  \\\ \delta_\xi^{(\eta)}  & \delta_\eta^{(\eta)} \end{array}\right).
\end{eqnarray}

The Frobenius norm of the Jacobian matrix, $\| \partial\vec{x}/\partial \vec{x}_0\|$ is the square root of the sum of the squares of the matrix elements and is also equal to the square root of the sum of the squares of the singular values of the Singular Value Decomposition (SVD) of the matrix,
\begin{eqnarray}
\left\| \frac{\partial\vec{x}}{\partial \vec{x}_0} \right\| &\equiv& \sqrt{ \left( \delta_\xi^{(\xi)} \right)^2+ \left( \delta_\xi^{(\eta} \right)^2+ \left( \delta_\eta^{(\xi)} \right)^2 +\ \left( \delta_\eta^{(\eta)} \right)^2}\label{Frob-exp}\nonumber\\\\ 
 & =& \sqrt{\Lambda_u^2+\Lambda_s^2}.
\end{eqnarray}  A $2\times2$ matrix has two singular values $\Lambda_u$ and $\Lambda_s$; by definition $\Lambda_u\geq\Lambda_s$.

When the flow is chaotic, neighboring streamlines separate exponentially. $\Lambda_u$ becomes exponentially large, and $\Lambda_s$ becomes exponentially small.  Consequently, $\Lambda_u$ is essentially equal to the Frobenius norm of the Jacobian matrix, $\| \partial\vec{x}/\partial \vec{x}_0\|$.  A full SVD analysis gives additional information, the directions in both $\xi_0,\eta_0$ space and in $\xi,\eta$ space in which trajectories exponentiate apart and exponentiate together. 

The natural logarithm of the Frobenius norm of the Jacobian matrix can be used to define the magnitude of the exponentiation when the exponentiation is large,
\begin{eqnarray}
&&\sigma(\vec{x}_0,\varphi)\nonumber\\
&&\equiv \ln\left(\sqrt{ \left( \delta_\xi^{(\xi)} \right)^2+ \left( \delta_\xi^{(\eta} \right)^2+ \left( \delta_\eta^{(\xi)} \right)^2 +\ \left( \delta_\eta^{(\eta)} \right)^2}\right). \label{sigma-def} \nonumber\\ 
\end{eqnarray}
The Frobenius norm  involves a sum of positive numbers and is far less numerically demanding than calculating the SVD or the Jacobian, which is the difference between two numbers, each of order the Frobenius norm squared.


\subsection{Chaos and disruptions}

In an evolving magnetic field, the magnitude of the exponentiation, $\sigma(\vec{x}_0,\varphi,t)$, depends on time.  As shown in \cite{Boozer:RMP}, an ideal evolution can be defined as being consistent with a time independent field line Hamiltonian $\psi_p(\psi,\theta,\varphi)$.  The changes in $\sigma$ come from the motion of the field lines.  The magnetic Reynold number $R_M$ is the ratio of the resistive $\mu_0a^2/\eta$ to the ideal-evolution $a/u$ timescale. When $\sigma$ becomes large compared to $R_M$ field line connections cannot be preserved \cite{Boozer:B-ev}.  An ideal evolution can take simple and smooth magnetic surfaces into a state in which the distance between neighboring magnetic surfaces varies by more than the magnetic Reynold number.  A rapid breaking of magnetic surfaces ensues, as in a tokamak disruption \cite{Boozer:surfaces,Jardin:2022}.  But, it should be noted that a chaotic flow will not reassemble the surfaces.  A tokamak disruption removes the drive for a non-axisymmetric equilibrium.  A disruption is so fast that the normal field to the wall remains axisymmetric, so an axisymmetric equilibrium with nested magnetic surfaces should re-form on a resistive timescale, not on a timescale determined by a near-ideal evolution that produces a large exponentiation $\sigma$.


\subsection{Chaos in divertors}

A steady-state divertor is characterized by a time independent magnetic field, which has a fixed $\vec{x}(\psi,\theta,\varphi)$ and Hamiltonian $\psi_p(\psi,\theta,\varphi)$.  For this problem, the position vector $\vec{x}(\psi,\theta,\varphi)$ is not the natural cause of a large exponentiation $\sigma$; the Hamiltonian is.  The chaos of the Hamiltonian can be defined by taking the position vector to be as trivial as possible, a periodic cylinder with a radial coordinate $r =\sqrt{\psi/\pi B}$ and a polar angle $\theta$.  The axial coordinate is $z\equiv R_0\varphi$, and all Fourier terms are consistent with the period of the cylinder, $2\pi R_0$.


\subsection{Hamiltonian produced chaos}

The cylindrical position vector is
\begin{eqnarray}
\vec{x} &=& R_0\varphi \hat{\varphi} + r \hat{r}(\theta),   \hspace{0.2in} \mbox{where} \\
\frac{d\hat{r}}{d\theta} &=& \hat{\theta} \hspace{0.2in}\mbox{and}\hspace{0.2in} \frac{d\hat{\theta}}{d\theta} = - \hat{r}.\\
\vec{B}&=& B_\varphi \hat{\varphi} + \frac{1}{2\pi R_0} \hat{\varphi}\times\vec{\nabla}\psi_p(r,\theta,\varphi)\\
\vec{h}&=& R_0\hat{\varphi} + \frac{1}{2\pi B}\Big( \frac{\partial\psi_p}{\partial r} \hat{\theta} -  \frac{1}{r} \frac{\partial\psi_p}{\partial \theta} \hat{r}\Big)
\end{eqnarray}

The two displacements are $\vec{\delta}_r\equiv (\delta r) \hat{r}$ and $\vec{\delta}_\theta \equiv (r\delta\theta) \hat{\theta}$.
\begin{eqnarray}
\vec{\delta}\cdot\vec{\nabla}\vec{h}&=& \frac{\delta_r}{2\pi B} \Big\{- \frac{\partial}{\partial r}\left(\frac{1}{r} \frac{\partial \psi_p}{\partial \theta}\right)\hat{r} +\frac{\partial^2\psi_p}{\partial r^2}\hat{\theta}\Big\} \nonumber\\
&&+\frac{\delta_\theta}{2\pi B} \Big\{- \left(\frac{1}{r^2}\frac{\partial^2\psi_p}{\partial\theta^2}+ \frac{1}{r}\frac{\partial \psi_p}{\partial r}\right)\hat{r} \nonumber\\
&& \left( -\frac{1}{r^2}\frac{\partial \psi_p}{\partial\theta} + \frac{1}{r}\frac{\partial^2 \psi_p}{\partial r\partial\theta}\right)\hat{\theta} \Big\},  
\end{eqnarray}
which implies
\begin{eqnarray}
\frac{d \delta_r}{d\varphi} &=& \delta_r D_{rr} + \delta_\theta D_{\theta r} \\
\frac{d \delta_\theta}{d\varphi} &=& \delta_r D_{r\theta} + \delta_\theta D_{\theta \theta} \\
D_{rr} &\equiv&- \frac{1}{2\pi B} \frac{\partial}{\partial r}\left(\frac{1}{r} \frac{\partial \psi_p}{\partial \theta}\right)\\
&=& -\frac{\partial^2 \psi_p}{\partial \psi\partial\theta} + \frac{1}{2\psi}\frac{\partial \psi_p}{\partial\theta}\\
D_{\theta r} &\equiv& - \frac{1}{2\pi Br^2}\frac{\partial^2\psi_p}{\partial\theta^2}+ \frac{1}{2\pi Br}\frac{\partial \psi_p}{\partial r} \\
&=&-\frac{1}{2\psi}\frac{\partial^2\psi_p}{\partial\theta^2} +  \frac{\partial \psi_p}{\partial \psi} \\
D_{r\theta} &\equiv& \frac{1}{2\pi B}\frac{\partial^2 \psi_p}{\partial r^2}\\
&=&2\psi \frac{\partial^2\psi_p}{\partial \psi^2} + \frac{\partial \psi_p}{\partial \psi}\\
D_{\theta \theta} &\equiv&  -\frac{1}{2\pi Br^2}\frac{\partial \psi_p}{\partial\theta} + \frac{1}{2\pi Br}\frac{\partial^2 \psi_p}{\partial r\partial\theta} \\
&=& \frac{\partial^2 \psi_p}{\partial \psi\partial\theta} -\frac{1}{2\psi}\frac{\partial \psi_p}{\partial\theta}
\end{eqnarray}


 \section{Effect of $\hat{\psi}_p$ variation on $\vec{j}$ \label{shape-j}}
 
 Although canonical transformations do not change the magnetic field, $\vec{B}(\vec{x})$, as a function of position, changes in the resonant Fourier terms change $\vec{B}(\vec{x})$ and hence the current density $\vec{j}(\vec{x})=\vec{\nabla}\times\vec{B}$.  Changes in the current density can be nulled or modified by small changes in the function $\vec{x}(\psi,\theta,\varphi)$,
 \begin{eqnarray}
2\pi\vec{B} =&& \vec{\nabla}\psi\times\vec{\nabla}\theta + \iota(\psi)\vec{\nabla}\varphi\times\vec{\nabla}\psi \nonumber\\ 
&&+\frac{\partial\hat{\psi}_p}{\partial\psi}\vec{\nabla}\varphi\times\vec{\nabla}\psi - \frac{\partial\hat{\psi}_p}{\partial\theta} \vec{\nabla}\theta \times\vec{\nabla}\varphi \\
=&& \frac{1}{\mathcal{J}}\Big(\frac{\partial \vec{x}}{\partial\varphi} +\iota \frac{\partial \vec{x}}{\partial\theta} +\frac{\partial\hat{\psi}_p}{\partial\psi}\frac{\partial \vec{x}}{\partial\theta} - \frac{\partial\hat{\psi}_p}{\partial\theta}\frac{\partial \vec{x}}{\partial\psi} \Big) \hspace{0.2in}\\
\mathcal{J}\equiv&& \left(\frac{\partial \vec{x}}{\partial\psi}\times \frac{\partial \vec{x}}{\partial\theta}\right)\cdot \frac{\partial \vec{x}}{\partial\varphi}.
\end{eqnarray} 

The current density is calculated using the covariant form for the magnetic field,
\begin{eqnarray}
&& \frac{2\pi}{\mu_0}\vec{B}= G_\varphi \vec{\nabla}\varphi + G_\theta \vec{\nabla}\theta + G_\psi \vec{\nabla}\psi.  \mbox{    Let   }\\
&& \gamma_{\psi\theta} \equiv \frac{ \frac{\partial\vec{x}}{\partial\psi} \cdot \frac{\partial\vec{x}}{\partial \theta}}{\mathcal{J}}, \mbox{ etc.,  }
\end{eqnarray}
then the covariant coefficients are
\begin{eqnarray}
&& G_\varphi = \frac{2\pi}{\mu_0} \vec{B}\cdot\frac{\partial\vec{x}}{\partial\varphi}\\
&&= \frac{2\pi}{\mu_0}\left\{\gamma_{\varphi\varphi} + \left(\iota +\frac{\partial\hat{\psi}_p}{\partial\psi}\right) \gamma_{\theta\varphi} -\frac{\partial\hat{\psi}_p}{\partial\theta} \gamma_{\psi\varphi} \right\}\\
&& G_\theta = \frac{2\pi}{\mu_0} \vec{B}\cdot\frac{\partial\vec{x}}{\partial\theta}\\
&&= \frac{2\pi}{\mu_0}\left\{\gamma_{\theta\varphi} + \left(\iota +\frac{\partial\hat{\psi}_p}{\partial\psi}\right) \gamma_{\theta\theta} -\frac{\partial\hat{\psi}_p}{\partial\theta} \gamma_{\psi\theta} \right\}\\
&& G_\psi = \frac{2\pi}{\mu_0} \vec{B}\cdot\frac{\partial\vec{x}}{\partial\psi}\\
&&= \frac{2\pi}{\mu_0}\left\{\gamma_{\psi\varphi} + \left(\iota +\frac{\partial\hat{\psi}_p}{\partial\psi}\right) \gamma_{\psi\theta} -\frac{\partial\hat{\psi}_p}{\partial\theta} \gamma_{\psi\psi} \right\}.\hspace{0.15in}
\end{eqnarray}
The current density 
\begin{eqnarray}
\vec{j} &=& \left(\frac{\partial G_\varphi}{\partial\theta} - \frac{\partial G_\theta}{\partial\varphi}\right)\frac{\frac{\partial \vec{x}}{\partial\psi}}{2\pi \mathcal{J}} + \left(\frac{\partial G_\psi}{\partial\varphi} - \frac{\partial G_\varphi}{\partial\psi}\right)\frac{\frac{\partial \vec{x}}{\partial\theta}}{2\pi \mathcal{J}} \nonumber\\
&& + \left(\frac{\partial G_\theta}{\partial\psi} - \frac{\partial G_\psi}{\partial\theta}\right)\frac{\frac{\partial \vec{x}}{\partial\varphi}}{2\pi \mathcal{J}}
\end{eqnarray}

There is an additional degree of freedom in the expression for the currents, the $G'$s.  This freedom is in the toroidal angle.  Instead, of the angle $\varphi$ of $(R,\varphi,Z)$ cylindrical coordinates, an angle $\zeta=\varphi +\omega(\psi,\theta,\zeta)$ can be used, where $\omega(\psi,\theta,\zeta)$ is any well-behaved function that is periodic in $\theta$ and $\zeta$.  When this change is made, $\vec{\nabla}\psi$ and $\vec{\nabla}\theta$ are unchanged, but $\vec{\nabla}\varphi$ becomes 
\begin{equation}
\vec{\nabla}\varphi = \left(1+\frac{\partial \omega}{\partial\zeta}\right) \vec{\nabla}\zeta + \frac{\partial \omega}{\partial\theta} \vec{\nabla}\theta + \frac{\partial \omega}{\partial\psi} \vec{\nabla}\psi.
\end{equation}

This change does not modify the enclosed poloidal current $\oint \vec{B}\cdot(\partial \vec{x}/\partial\varphi)d\varphi/\mu_0$ or the enclosed toroidal current $\oint \vec{B}\cdot(\partial \vec{x}/\partial\theta)d\theta/\mu_0$ but does change the current density at a given point in $(\psi,\theta,\zeta)$ space when the definition of $\zeta$ is changed.


\end{document}